# Experimental Demonstration of an Extreme Sub-Wavelength Nanomagnetic Acoustic Antenna Actuated by Spin-Orbit Torque from a Heavy Metal Nanostrip

Dr. Md Ahsanul Abeed and Prof. Supriyo Bandyopadhyay[*]

Department of Electrical and Computer Engineering

Virginia Commonwealth University, Richmond, VA 23284, USA

E-mail: sbandy@vcu.edu

A novel on-chip extreme sub-wavelength "acoustic antenna" whose radiation efficiency is ~50 times larger than the theoretical limit for a resonantly driven antenna is demonstrated. The antenna is composed of magnetostrictive nanomagnets deposited on a piezoelectric substrate. The nanomagnets are partially in contact with a heavy metal (Pt) nanostrip. Passage of alternating current through the nanostrip exerts alternating spin-orbit torque on the nanomagnets and periodically rotates their magnetizations. During the rotation, the magnetostrictive nanomagnets expand and contract, thereby setting up alternating tensile and compressive strain in the piezoelectric substrate underneath. This leads to the generation of a surface acoustic wave in the substrate and makes the nanomagnet assembly act as an acoustic antenna. The measured radiation efficiency of this acoustic antenna at the detected frequency is ~1%, while the wavelength to antenna dimension ratio is ~ 67:1. For a standard antenna driven at acoustic resonance, the efficiency would have been limited to ~ $(1/67)^2$ = 0.02%. It was possible to beat that limit (by ~50 times) via actuating the antenna not by acoustic resonance, but by using a completely different mechanism involving spin-orbit torque originating from the giant spin Hall effect in Pt.

# I. INTRODUCTION

There is a strong interest in sub-wavelength antennas of all types whose physical dimensions are much smaller than the wavelength of the radiation they emit. This allows them to be integrated into embedded systems (e.g. wearable electronics) or in medically implanted devices or in miniaturized communicators. However, such antennas typically have very poor radiation efficiency when they are excited at the radiation resonance (an electromagnetic antenna excited by an electromagnetic wave at electromagnetic resonance, or an acoustic antenna excited by an acoustic wave at acoustic resonance, etc.), because the efficiency is $\sim (l/\lambda)^2$ $[l < \lambda]$, where $l$ is the antenna dimension and $\lambda$ is the wavelength of the emitted radiation.

Recently, some work has been reported in driving an electromagnetic antenna not with an electromagnetic wave, but with an acoustic wave at *acoustic resonance* [1-5]. The idea is to generate electromagnetic waves by periodically flipping the magnetization of magnetostrictive nanomagnets with periodic strain caused by an acoustic wave launched in a piezoelectric substrate underneath the nanomagnets. The oscillating magnetizations emanate an electromagnetic wave at the same frequency as the acoustic wave. The system therefore acts as an electromagnetic antenna driven at acoustic resonance. Since at the same frequency, the wavelength of an acoustic wave in many piezoelectric substrates is about five orders of magnitude smaller than that of an electromagnetic wave, the ratio $(l/\lambda)^2$ can be ten orders of magnitude larger if the electromagnetic antenna is excited by an acoustic wave instead of an electromagnetic wave. This principle allows one to build an extreme sub-wavelength electromagnetic antenna with reasonable efficiency. To put this into perspective, consider the fact that in the high frequency band (3 MHz – 30 MHz), the wavelength of electromagnetic wave is 10 m to 100 m. A traditional electromagnetic antenna driven at electromagnetic resonance and radiating in this frequency range will have to have a size at least 1 – 10 m to be sufficiently efficient. However, an acoustically driven nanomagnetic antenna of the above type can have a size of few μm and yet be just as efficient! This is the principle to implement extreme sub-wavelength antennas.

In this work, we demonstrate an extreme sub-wavelength acoustic antenna that emits acoustic waves (in addition to electromagnetic waves) with an efficiency of ~1%. The antenna dimension is 67 times smaller than the acoustic wavelength, so the efficiency would have been limited to $(1/67)^2 = 0.02\%$ if we had driven the antenna at acoustic resonance. To overcome that limit and achieve an efficiency 50 times larger than the limit, we adopted a different principle of actuation. An alternating charge current, passed through a heavy metal (Pt) strip that is in contact with an array of nanomagnets, produces alternating spin-orbit torque on the nanomagnets because of the giant spin Hall effect in Pt. As long as the period of the alternating charge current exceeds the time required to rotate the magnetization of the nanomagnets through a significant angle, the magnetizations of the nanomagnets will rotate periodically with sufficient amplitude (or flip periodically) and emit an electromagnetic wave. At the same time, if the nanomagnets are magnetostrictive, then they will periodically expand and contract when their magnetizations are rotating, assuming *they are not clamped by the Pt strip*. If the nanomagnets are deposited on a piezoelectric substrate, then their periodic expansion/contraction will generate a periodic strain in the underlying piezoelectric, leading to the propagation of a surface acoustic wave (SAW) in the substrate that can be detected as an oscillating electrical signal with interdigitated transducers (IDT). This is the principle of the acoustic antenna. The wavelength of the acoustic wave is determined solely by the frequency of the alternating charge current (which is the dominant frequency of the generated acoustic wave) and the velocity of acoustic wave propagation in the piezoelectric substrate. Therefore, it has no relation to the size of the nanomagnets (antenna elements) which can be much smaller than the size of the acoustic wavelength. Thus, this construct can be a sub-wavelength acoustic antenna with a radiation efficiency that exceeds the theoretical limit for an acoustic antenna that is excited at acoustic resonance.

Acoustic antennas with enhanced radiation efficiencies have been studied earlier. Ref. [6] studied an acoustic antenna predicated on the acoustic Purcell effect and it requires a cavity that modifies the acoustic density of states. This is somewhat unrelated to our work. Ref. [7] studied a sub-wavelength acoustic antenna that uses Mie resonance and observed a 2.33 fold increase in the radiation intensity, while ref. [8] studied a 10-fold increase in the directivity of an acoustic antenna. Our antenna is omnidirectional in the plane of the piezoelectric substrate and the increase in efficiency/radiation intensity over that of a standard acoustically excited antenna is 50-fold.

The proposed acoustic antenna is shown schematically in Fig. 1. The nanomagnets are fabricated with a ledge (as shown in the inset) which is placed underneath a heavy metal (Pt) nanostrip. The bulk of the nanomagnet is outside the strip and hence its expansion/contraction is *not clamped* by the nanostrip. Only the ledge is clamped.

When a charge current is injected into the nanostrip, the top and bottom surfaces of the nanostrip become spin-polarized because of the giant spin Hall effect in Pt. The two surfaces have antiparallel polarizations. The polarizations of spins in either surface will depend on the direction of the current and will change sign when the current reverses direction. The accumulated spins in the *bottom* surface of the nanostrip will diffuse into the "ledges" that they are in contact with, and from there into the nanomagnets, which will exert a spin-orbit torque on the latter and rotate their magnetizations. If we reverse the direction of the injected charge current, then that will reverse the spin polarization of the bottom surface of the nanostrip and hence rotate the magnetizations of the nanomagnets in the opposite direction because the spin-orbit torque will reverse direction. This will happen only as long as the period of the current is longer than the time of magnetic reversal. Thus, if we pass an alternating current through the nanostrip, we will rotate or flip the magnetizations of the nanomagnets periodically, as long as the frequency of the current is considerably smaller than the inverse of the spin flip times of the nanomagnets. This alternating rotation will emit an electromagnetic wave, not unlike spin torque nano-oscillators [9, 10], and hence the system will act as an electromagnetic antenna. At the same time, the nanomagnets will periodically expand and contract because they are magnetostrictive, thus executing a "breathing mode" oscillation. This will generate periodic strain in the piezoelectric substrate underneath the nanomagnets and set up a SAW that can be detected with IDTs. Thus, this system acts as a dual electromagnetic and acoustic antenna. Here, we have investigated only the acoustic antenna functionality.

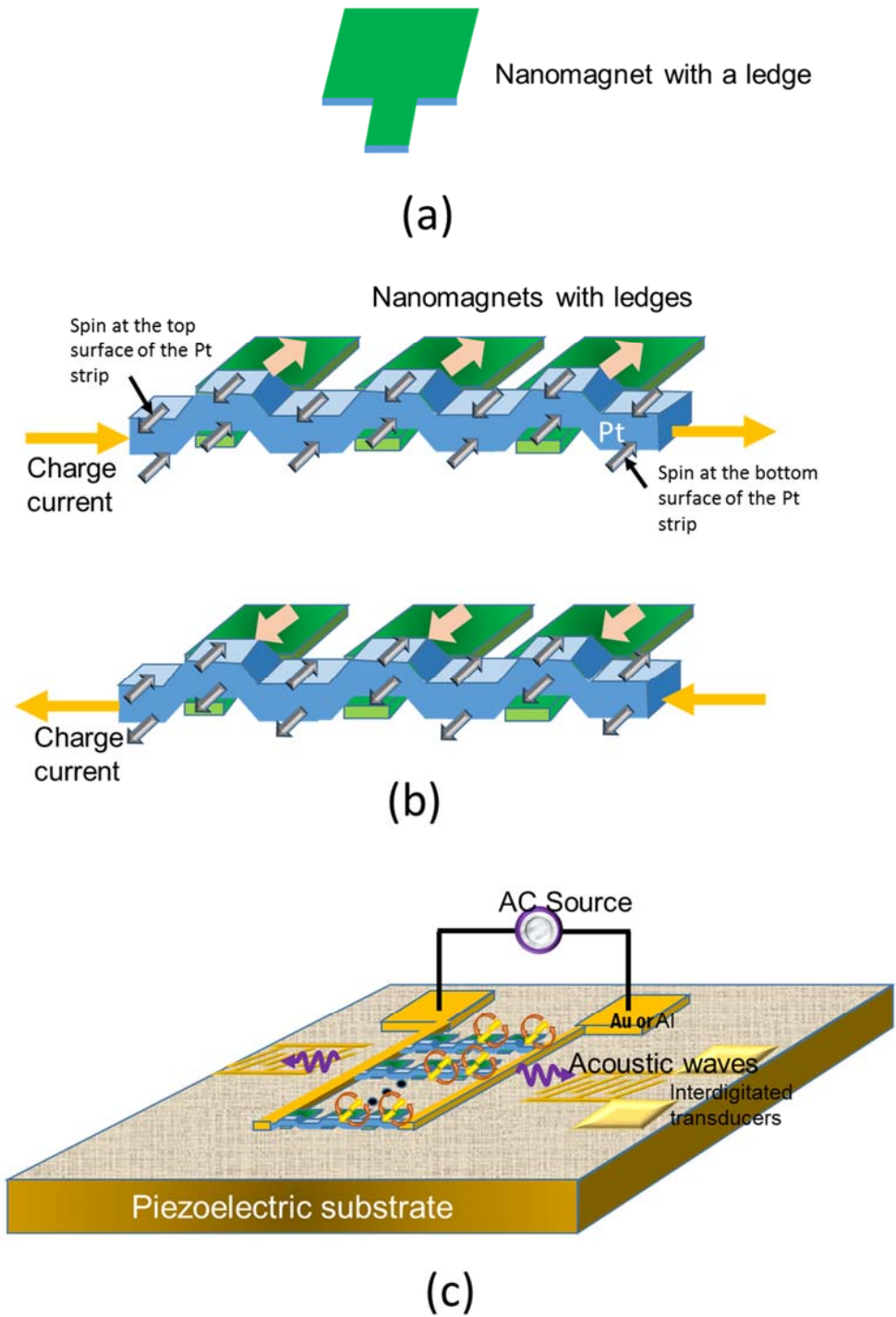

Fig. 1: (a) A rectangular nanomagnet with a ledge. (b) Principle of actuation of the antenna by the spin orbit torque in the heavy metal (Pt) nanostrip. For one polarity of the injected charge current, the nanomagnets are magnetized in one direction and for the opposite polarity, they are magnetized in the opposite direction. (c) The nanomagnets expand or contract as they transition between the two orientations and that generates a surface acoustic wave in the piezoelectric substrate which can be detected by interdigitated transducers delineated on the substrate.

In some sense, this acoustic antenna is the *converse* of the sub-wavelength electromagnetic antenna reported in ref. [3]. There, strain (phonons) were used to generate electromagnetic waves (photons), while here electrical signal (photons) have been used to generate acoustic waves (phonons).

## II. RESULTS

In Fig. 2(a), we show the pattern of the acoustic antenna (with the nanomagnets, Pt lines, contact pads to the Pt lines and IDTs) and in Fig. 2(b), we show a scanning electron micrograph of the fabricated nanomagnets. The nanomagnets are rectangular with long dimension ~250 nm, short dimension ~200 nm and the ledge length is ~100 nm. The ledge has a "Gaussian" shape and the full width at half maximum is ~70 nm. In Fig. 3, we show scanning electron micrographs of the Pt line and nanomagnet assembly. There are 40 Pt lines and hence 40 rows of nanomagnets that are contacted.

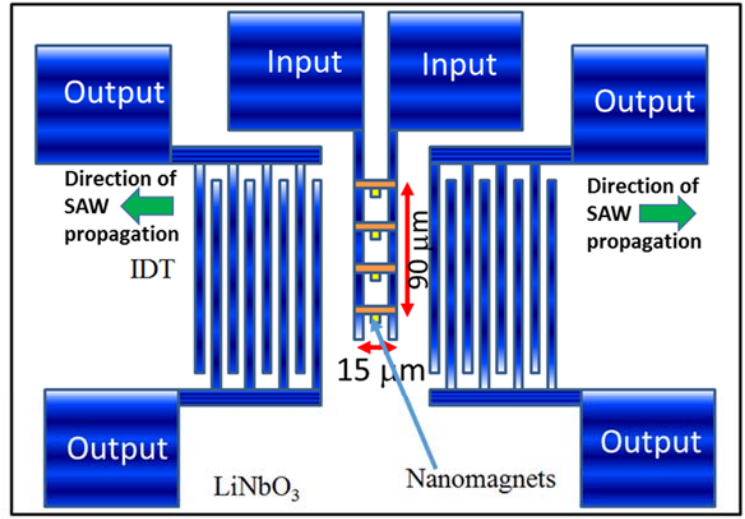

(a)

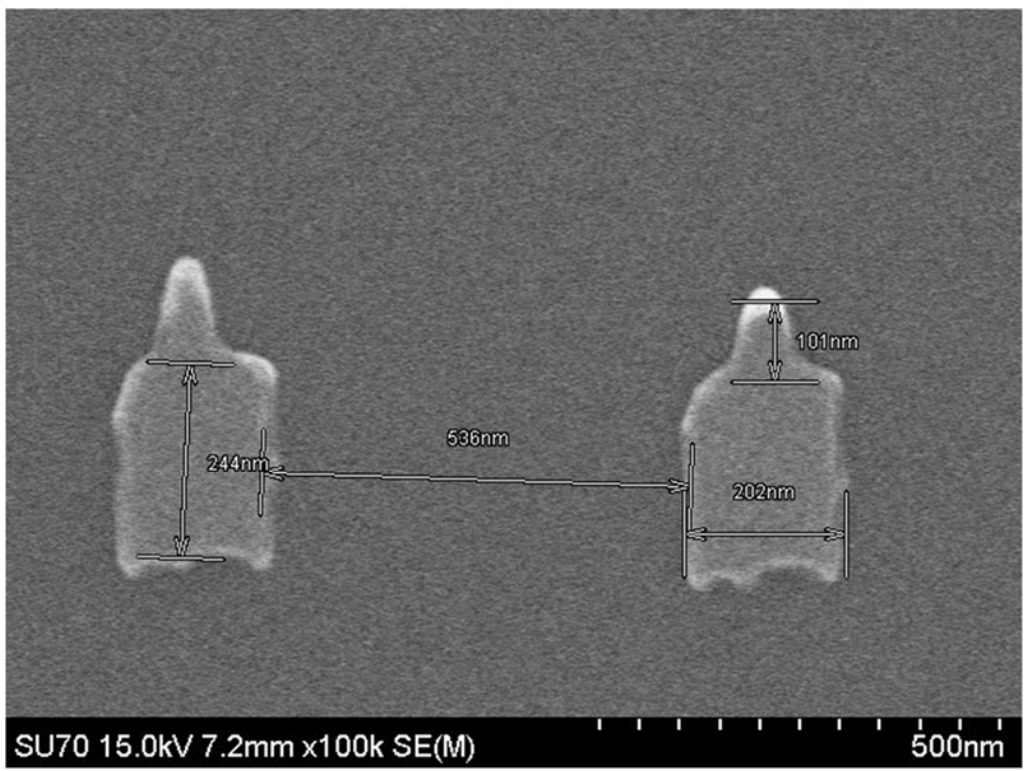

(b)

Fig. 2: (a) Pattern for the acoustic antenna. This figure is not to scale. (b) Scanning electron micrographs of the fabricated Co nanomagnets with Gaussian shaped ledges.

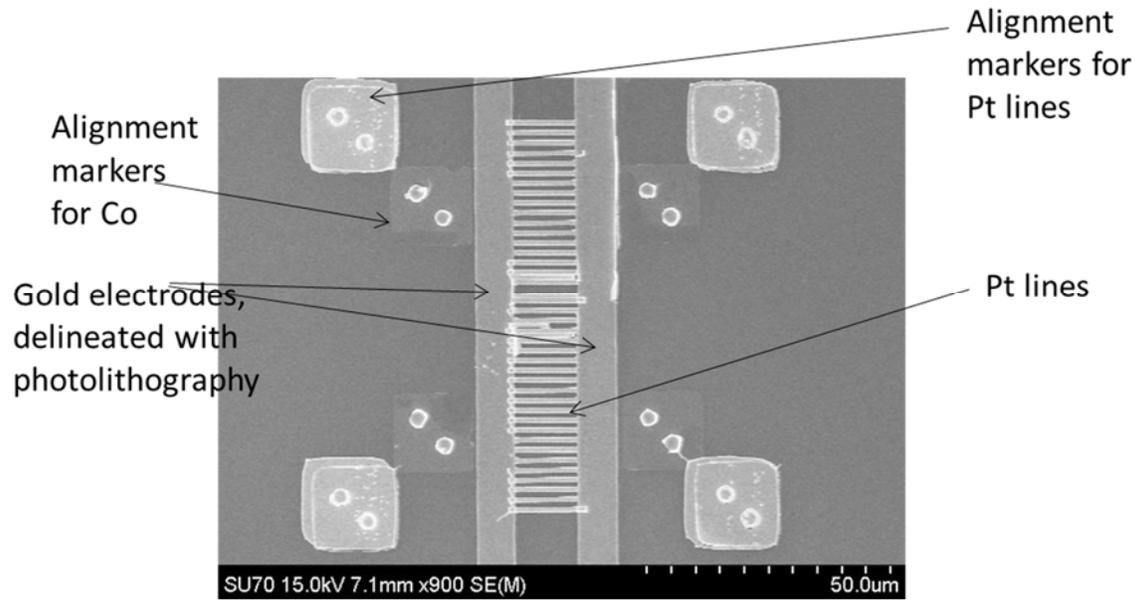

(a)

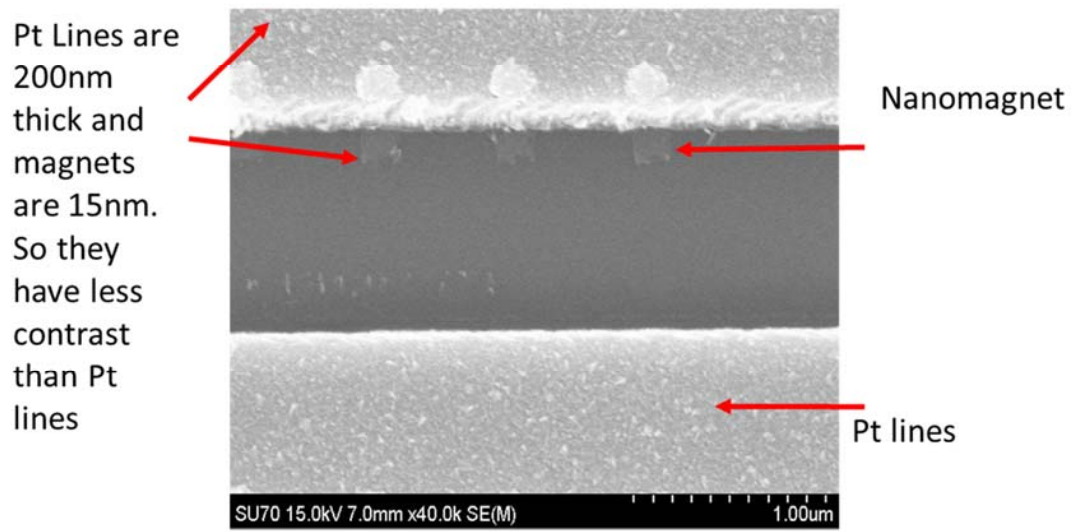

(b)

Fig. 3: (a) Scanning electron micrograph of the Pt lines overlying the nanomagnets. (b) Zoomed view showing the nanomagnets underneath the Pt line.

In Fig. 4, we show both the oscilloscope traces of the sinusoidal voltage applied across the Pt lines to actuate the acoustic antenna via spin-orbit torque, and the voltage detected at the interdigitated transducers (IDT). We show these traces for two frequencies, 3.63 MHz which is the resonant frequency of the IDT, and 6.87 MHz. Traces for a few other frequencies can be found in the Supporting Information.

The resistance of the 40 parallel Pt lines varied between 98 ohms and 108 ohms from sample to sample. Therefore, the resistance of each line is on the order of 4 k$\Omega$. For the case in Fig. 4(a), the input voltage of 11.25 V peak-to-zero [measured from the oscilloscope trace] will produce a current of 2.8 mA peak-to-zero in a Pt line. The line has a length of 15 μm, width 1 μm and thickness 200 nm. Therefore, the peak current density in each Pt line is $1.4 \times 10^{10}$ A m$^{-2}$, which should be well above the critical current density needed to produce sufficient spin-orbit torque. In the case of Fig. 4(b), the input voltage of 12.85 V peak-to-zero produced a peak-to-zero current of 3.21 mA in a Pt line, resulting in a peak-to-zero current density of $1.6 \times 10^{10}$ A m$^{-2}$ in each line.

The input power to the acoustic antenna is calculated as $V_{in}^2/2R_{Pt}$ where $V_{in}$ is the peak-to-zero input voltage and $R_{Pt}$ is the resistance of the 40 Pt lines in parallel. For the case in Fig. 4(a), this quantity is 633 mW, while for the case in Fig. 4(b), it is 825 mW.

In order to calculate the radiation efficiency, we have to know the power in the acoustic wave that has been produced. The power carried by an acoustic wave of amplitude $\phi$ is given by [11]

$$P = \frac{1}{2} y_0 \frac{W}{\lambda} \phi^2, \qquad (1)$$

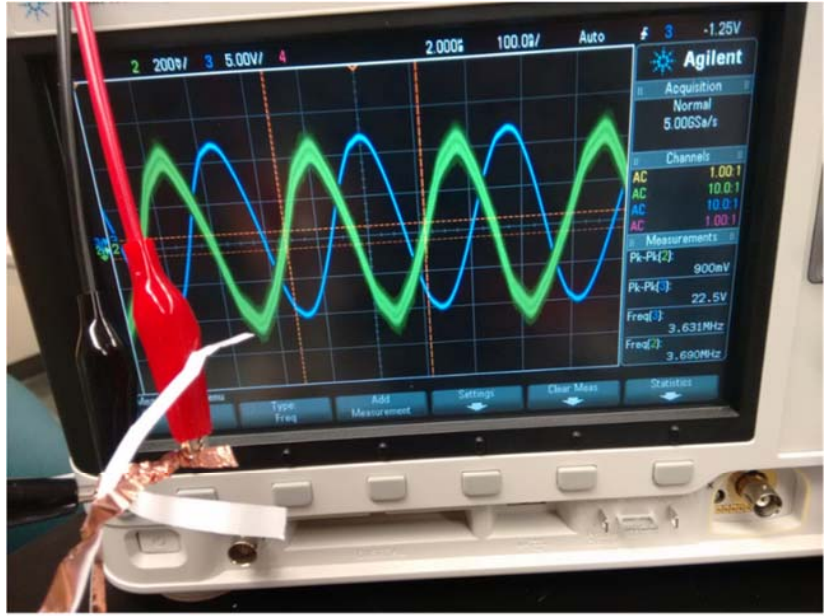

(a)

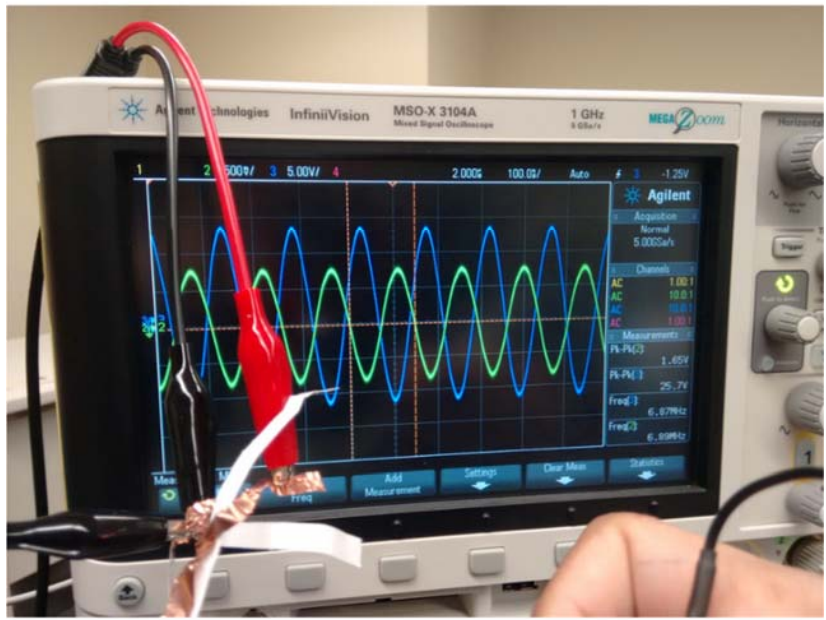

(b)

Fig. 4: Oscilloscope traces of the alternating voltage applied across the Pt lines to actuate the acoustic antenna (blue) and the alternating voltage detected at the interdigitated transducer (green). They are respectively the input and output signals. (a) The input voltage frequency is 3.63 MHz which is the resonant frequency of the interdigitated transducers (IDT) determined by the spacing of the IDT fingers and the velocity of surface acoustic wave in the substrate. Input voltage peak to peak amplitude is 22.5 V and the detected voltage peak-to-peak amplitude is 0.9 V. (b) The input voltage frequency is 6.87 MHz and the peak-to-peak amplitude is 25.7 V, while the detected voltage peak-to-peak amplitude is 1.65 V.

where $y_0$ is the characteristic admittance of the SAW line and has a value of $2.1 \times 10^{-4}$ S for LiNbO$_3$ [11], $W$ is the width of the IDT and $\lambda$ is the wavelength of the SAW. The IDTs were designed and fabricated for $W/\lambda = 40$.

Neglecting capacitive and inductive effects, the voltage $V_{out}$ detected at the IDT is related to the SAW amplitude $\phi$ as [11]

$$\phi \approx \mu V_{out}, \qquad (2)$$

where $\mu$ is the response function of an IDT operating in the transmitting mode. For our system, this quantity was calculated in ref. [12] as approximately 2. Hence, for the case shown in Fig. 4(a) [where $V_{out}$ = 0.45 V peak-to-zero as seen in the oscilloscope trace] $\phi$ = 0.9 V peak-to-zero and for that shown in Fig. 4(b), $\phi$ = 1.6 V peak-to-zero. Therefore, from Equation (2), the SAW power produced in the two cases are 3.4 mW and 10.7 mW, respectively. The corresponding efficiencies of SAW production are 0.54% and 1.3%, respectively. These numbers are approximate because of some simplifying assumptions such as neglecting inductive and capacitive effects. We have also assumed that the IDT detection efficiency is 100%, which is an over-estimate, especially at frequencies that are not the IDT resonant frequency. Hence, the estimates of the antenna efficiencies presented here are conservative.

In our LiNbO$_3$ substrate, the acoustic wave velocity is ~ 3300 m/s. Therefore, for a frequency of 3.63 MHz, the acoustic wavelength is ~ 1 mm. The nanomagnet assembly acting as the antenna has a dimension of ~ 15 μm in the direction of SAW propagation. Hence the *ratio of acoustic wavelength to antenna dimension* is ~67, making it an extreme *sub-wavelength* antenna. If the antenna was excited by an acoustic wave and driven at the acoustic resonance, the radiation efficiency would have been limited to ~$(1/67)^2$ = 0.02%. Our efficiency is about 50 times larger.

We were able to overcome the limit because we actuated the antenna via spin-orbit torque and did not drive it with an acoustic wave.

One important question that needs to be resolved before proceeding further is whether the output voltage detected at the IDT (shown in Fig. 4) could have been due to direct electromagnetic pick up through the air from the input, instead of being from an acoustic wave in the substrate. Electromagnetic pick up, however, could never have produced the phase shifts (time delay) seen between the input and output signals in Fig. 4. The measured average distance between the input and output ports is ~6 mm and the time $\Delta t$ that it would take an electromagnetic wave to traverse this distance is 20 ps. At a frequency $f$ of 3.63 MHz (Fig. 4a), this would produce a phase lag of $2\pi f \Delta t = 4.56 \times 10^{-4}$ radians between the output and input, which is much smaller than what is observed. For an acoustic wave traveling in the substrate, with a velocity 5 orders of magnitude smaller than that of an electromagnetic wave propagating through the air, $\Delta t$ will be five orders of magnitude larger and the phase shift at this frequency will be 45.6 radians = $(14\pi + 1.62)$ radians. When the modulo $2\pi$ value of this phase shift is taken, this is 1.62 radians. The observed phase shift is about 2.2 radians which is much closer to the acoustic phase shift than the electromagnetic phase shift. This gives us confidence that the detected signal is not due to electromagnetic pick up.

Repeating this exercise for the 6.87 MHz frequency will yield an electromagnetic phase shift of $8.63 \times 10^{-4}$ radians and an acoustic phase shift of 86.3 radians = $(26\pi + 4.6)$ radians. The modulo $2\pi$ value of this phase shift is 4.6 radians which is close to the observed value of 3.3 radians. This again gives us confidence that the observed output voltage at the IDT is indeed due to the generated surface acoustic wave.

Finally, it should be obvious that had we used the IDT as the input port and the two terminals of the Pt strip as output ports, then an oscillating electrical signal applied at the input port would have, in principle, also produced an oscillating electrical signal at the output port because of reciprocity [13]. The input signal at the IDT would have launched a SAW in the substrate, which would have periodically rotated the magnetizations of the magnetostrictive nanomagnets owing to the Villari effect. This would have resulted in periodic spin pumping into the Pt line, resulting in an alternating spin current which would have been converted into a charge current by the inverse spin Hall effect, resulting in an oscillating electrical signal across the Pt lines. However, to observe the reciprocal effect, the design should be such that sufficient spin pumping can occur. We intentionally designed the structure to work optimally as an acoustic antenna, which is why we designed the nanomagnets with "ledges" to avoid clamping. The flip side of this is that it precludes sufficient spin pumping because a small section of the nanomagnets are in contact with the Pt strip. Therefore, the reciprocal effect will be small in our structure and probably outside the detection limit of our equipment.

### III.  CONCLUSIONS

In conclusion, we have demonstrated an acoustic antenna actuated by the spin-orbit torque from a heavy metal nanostrip. The use of this novel actuation mechanism allowed us to build an *extreme sub-wavelength* acoustic antenna with a radiation efficiency over 50 times larger than the limit for an acoustic antenna actuated by an acoustic wave and operated at acoustic resonance. Antennas such as these have applications in many areas such as miniaturized speakers, micro electro-mechanical (MEMS) devices, acoustic mapping and analysis of biological specimens in a biochip or biosensor, etc.

From the physics perspective, this antenna converts photons in the input (low-frequency electromagnetic) alternating signal applied to the Pt strip to magnons via the spin Hall effect (or spin-orbit torque) and then to phonons in the surface acoustic wave via magneto-elastic coupling, which are finally detected at the IDT. The overall efficiency of this three-step process is ~1%. Ref. [13] reported a similar phenomenon of converting spin current to phonons, but not in the context of an antenna, although the physics is the same.

Interestingly, in many ways, the phenomenon we report for the acoustic antenna is the converse of that reported in ref. [3] for an electromagnetic antenna. The construct of ref. [3] converted phonons to magnons to photons in order to realize a sub-wavelength *electromagnetic* antenna implemented with magnetostrictive nanomagnets that were periodically strained with a surface acoustic wave. We, too, have been able to implement a reasonably efficient extreme sub-wavelength electromagnetic antenna in that fashion, but that outside the scope of this work and will be reported elsewhere.

Before we conclude, we discuss one other aspect. The excitation frequency (frequency of the electrical signal applied to the Pt nanostrip to produce the giant spin Hall effect) in this work was around 3.6 MHz. The maximum allowable frequency (or alternately the minimum signal period) is determined by the time it takes for the nanomagnets' magnetizations to rotate over a significantly large angle. The signal period must exceed the latter time by a factor of perhaps ~10 to ensure that nearly all the nanomagnets have ample time to rotate their magnetizations through a large enough angle and produce a strain in the piezoelectric substrate underneath. The time taken by the magnetization to rotate through a large enough angle depends on the size, shape, nanomagnet material, and also the strength of the spin orbit torque, which, in turn, is determined by the spin-orbit interaction strength in the heavy metal (in this case platinum) and the magnitude of the current injected into it. For reasonable values of these parameters, we estimate that the switching time is no less than ~1 ns. Our own past simulations based on the Landau-Lifshitz-Gilbert equation suggests that the switching time is on the order of 1 ns. Of course, there will be distribution of the switching time because of defects in the nanomagnets, pinning sites, variations in shape and size, etc, As a result, we would say that a safe estimate would be 10 ns, which would then limit the frequency to perhaps 100 MHz. This is enough for many on-chip acoustic applications.

## IV. EXPERIMENTAL SECTION

The nanomagnets and Pt strip were fabricated on a $128^0$ Y-cut $LiNbO_3$ substrate. The substrate was spin-coated with bilayer PMMA e-beam resists of different molecular weights to obtain good undercut: PMMA 495 diluted 4% by volume in Anisole, followed by PMMA 950 also diluted 4% by volume in Anisole. The spin coating was carried out at a spin rate of 2500 rpm. The resists were subsequently baked at $110^0$ Celsius for 5 min. Next, electron-beam lithography was performed using a Hitachi SU-70 scanning electron microscope (at an accelerating voltage of 30 kV and 60 pA beam current) with a Nabity NPGS lithography attachment. Finally, the resists were developed in MIBK−IPA (1:3) for 270 s followed by a cold IPA rinse.

For nanomagnet delineation, a 5 nm thick Ti adhesion layer was first deposited on the patterned substrate using e-beam evaporation at a base pressure of ~$2 \times 10^{-7}$ Torr, followed by the deposition of Co. Pt was deposited similarly. The lift-off was carried out using Remover PG solution.

**Supporting Information**

Supporting information is available from the Wiley Online Library or the corresponding author.

**ACKNOWLEDGMENTS**

This work was supported in part by the US National Science Foundation under grants ECCS-1609303 and CCF-1815033. It was also supported by a Virginia Commonwealth University Commercialization grant and an Indo-US Science and Technology Fund Center grant titled "Center for Nanomagnetics for Energy-Efficient Computing, Communications and Data Storage" (IUSSTTF/JC-30/2018). M. A. A. carried out all the fabrication. S. B. conceived of the idea. Both authors collected and analyzed data and contributed to writing the paper.

# Supporting Information: Experimental Demonstration of a Nanomagnetic Acoustic Antenna Actuated by Spin-Orbit Torque from a Heavy Metal Nanostrip

Md Ahsanul Abeed and Supriyo Bandyopadhyay

Department of Electrical and Computer Engineering

Virginia Commonwealth University, Richmond, VA 23284, USA

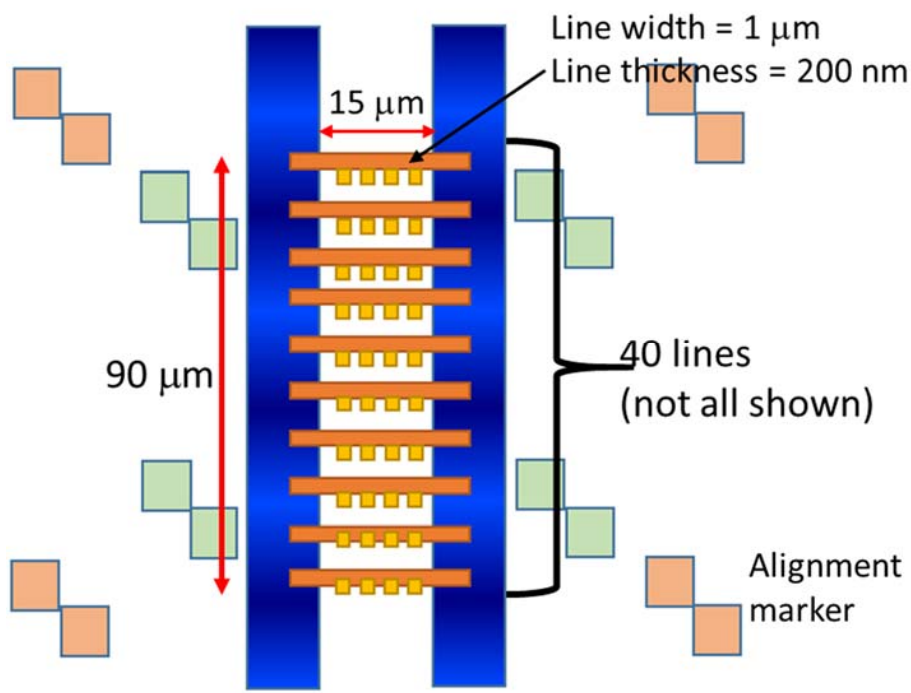

Fig. S1: The acoustic antenna pattern with alignment markers.

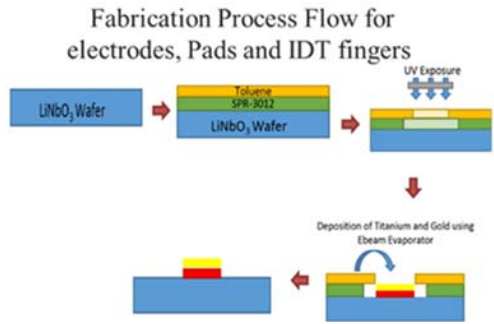
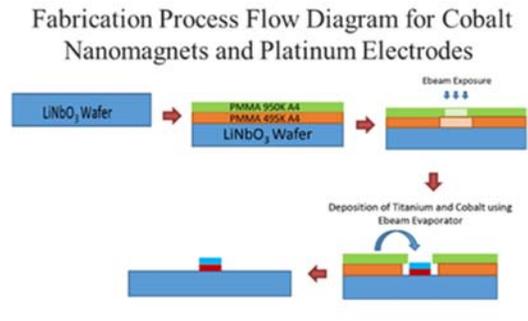
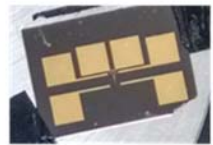
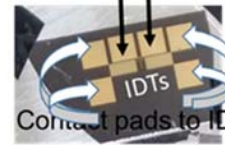

Fig. S2: Fabrication flow chart and top views of the fabricated devices. Design II had IDTs while Design I has just metal pads. Only Design II was used in the experiments.

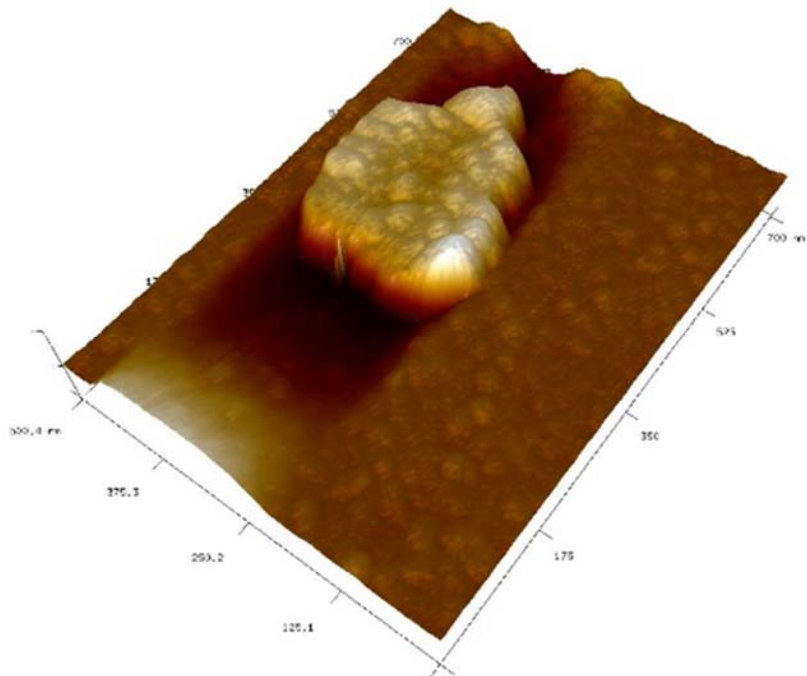

Fig. S3: Atomic force micrograph of a single nanomagnet with ledge.

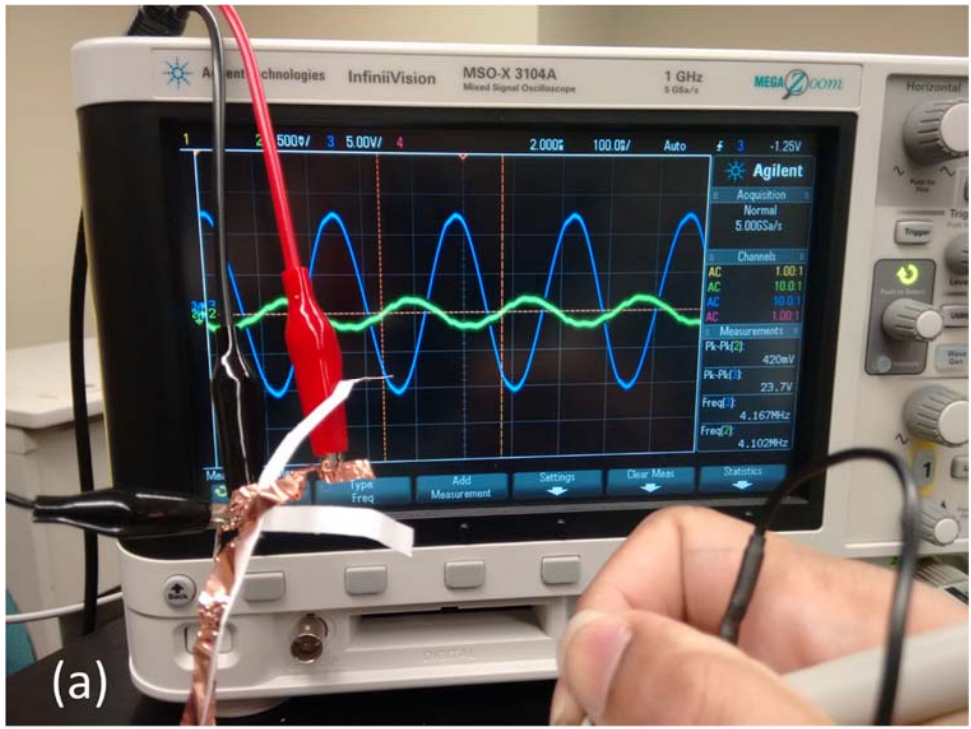

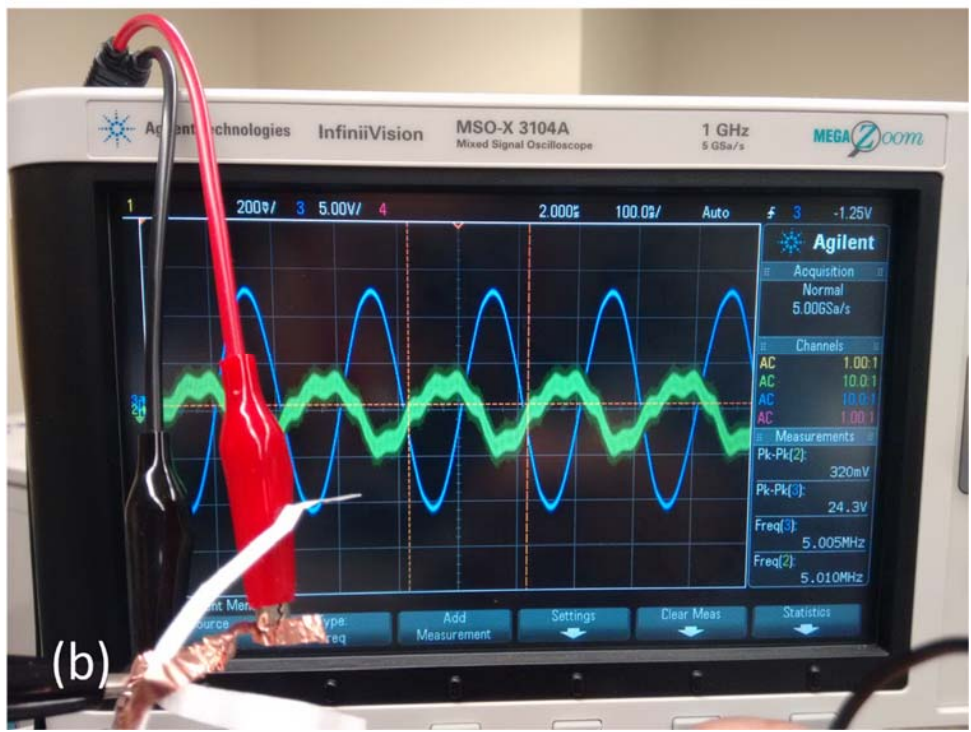

Fig. S4: Oscilloscope traces of input (blue) and output (green) signals at frequencies of (a) 4.167 MHz and (b) 5.005 MHz.

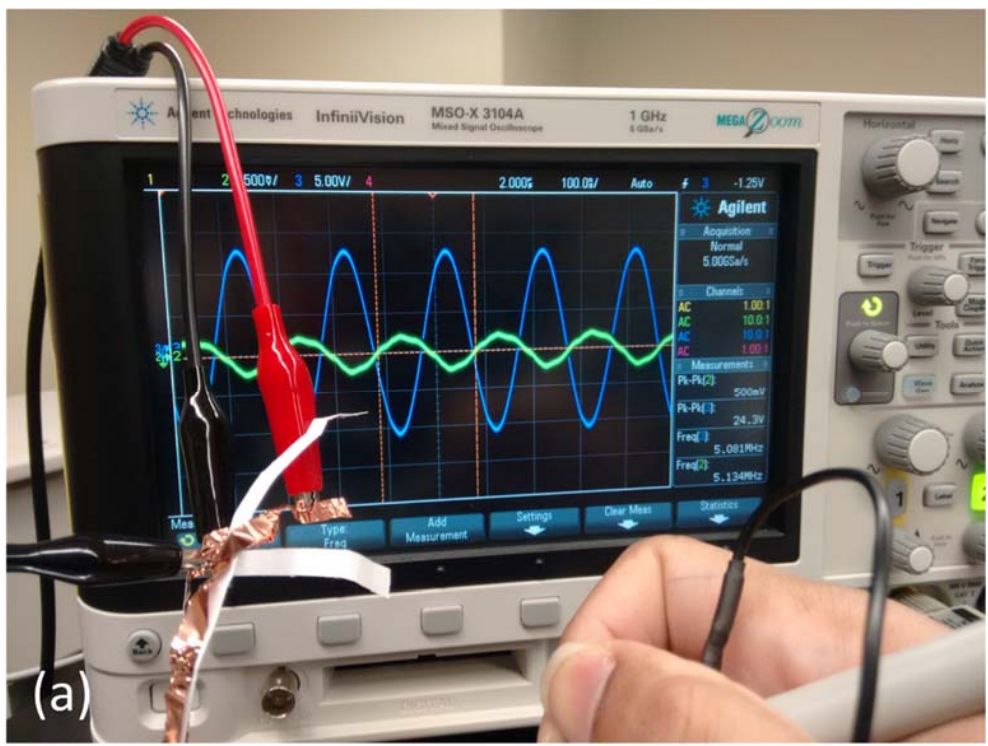

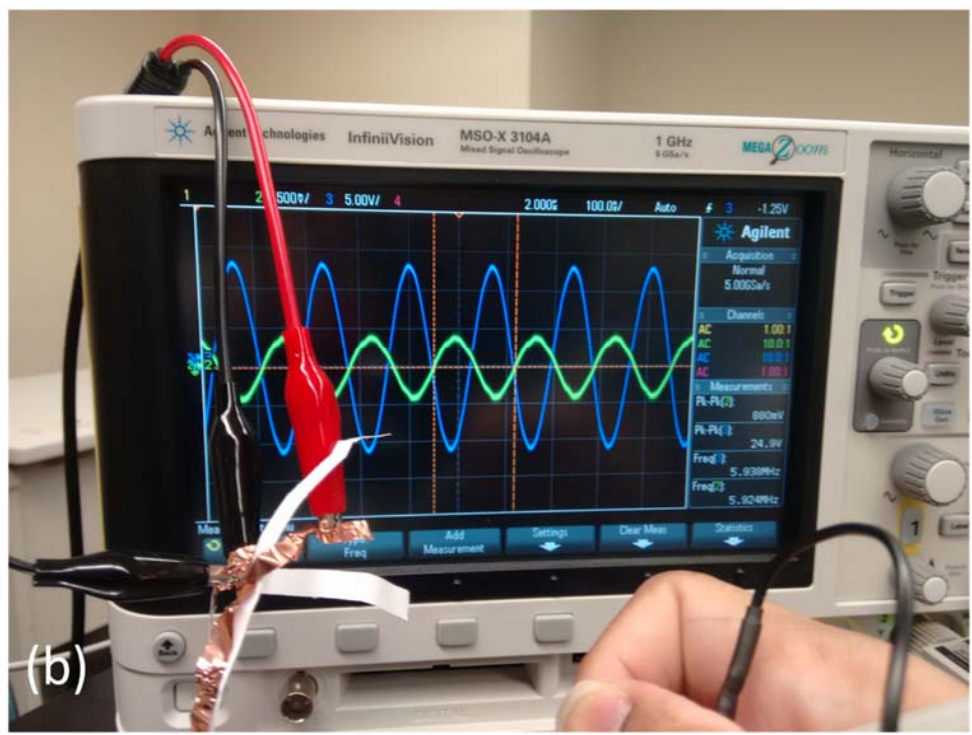

Fig. S5: Oscilloscope traces of input (blue) and output (green) signals at frequencies of (a) 5.08 MHz and (b) 5.938 MHz.

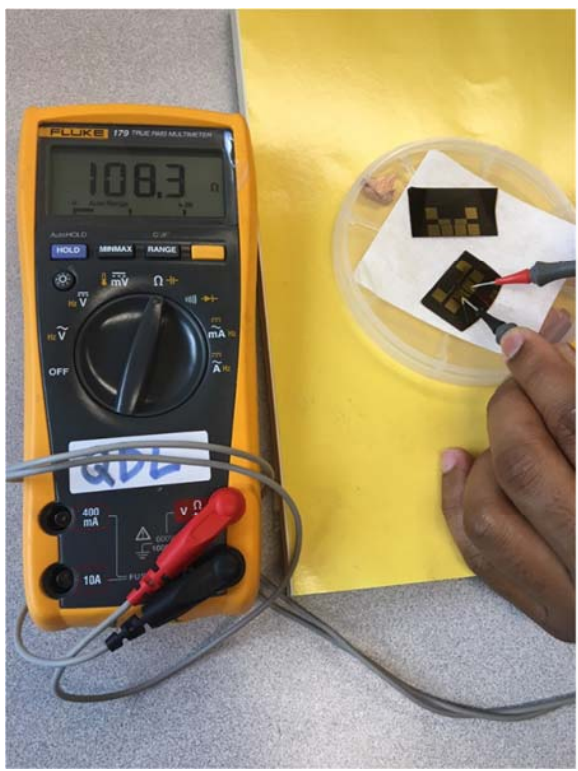 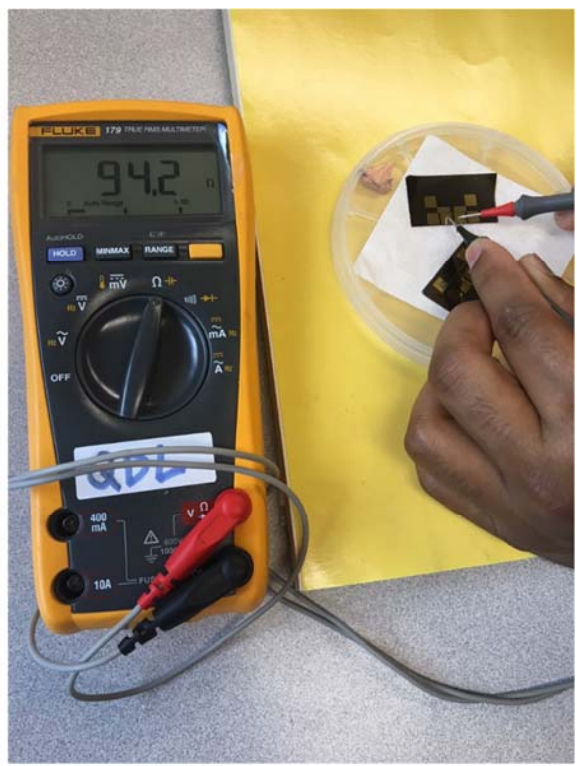

Fig. S6: The resistance of the Pt lines (two different samples).